\documentclass[onecolumn,aps,nofootinbib,notitlepage]{revtex4-1}

\usepackage[margin=2cm]{geometry}
\usepackage{amsfonts}
\usepackage{amsmath}
\usepackage{amsthm}
\usepackage{amssymb}
\usepackage[svgnames]{xcolor}
\usepackage[hidelinks]{hyperref}
\hypersetup{
    colorlinks=true,
    linkcolor=NavyBlue,
    filecolor=NavyBlue,
    urlcolor=NavyBlue,
    citecolor=NavyBlue,
    }

\usepackage{graphics}
\usepackage{tikz}
\usepackage{graphicx}
\usetikzlibrary{positioning}
\usepackage{caption}

\usepackage{subcaption}
\usepackage{adjustbox}
\usepackage{gensymb}
\usepackage{breqn}
\usepackage{braket}
\usepackage{enumitem}

\usepackage[compat=1.0.0]{tikz-feynman}

\def\gev{\, {\rm GeV}}
\def\mev{\, {\rm MeV}}

\def\be{\begin{equation}}
\def\ee{\end{equation}}
\def\bea{\begin{eqnarray}}
\def\eea{\end{eqnarray}}
\newcommand{\Dsl}[1]{\slash\hskip -0.20 cm #1}

\begin{document}

\begin{center}
\hfill  MI-TH-203, UH511-1310-2020
\end{center}

\title{\large{\textbf{Contributions to $\Delta N_{eff}$ from the dark photon of $U(1)_{T3R}$}}}
\author{\normalsize{ Bhaskar Dutta$^{\bf 1}$\footnote{dutta@physics.tamu.edu}, Sumit Ghosh$^{\bf 1}$\footnote{ghosh@tamu.edu}, Jason Kumar$^{\bf 2}$\footnote{jkumar@hawaii.edu}} \\
\vspace{1.0cm}
\normalsize\emph{$^{\bf 1}$Mitchell Institute for Fundamental Physics and Astronomy, Department of Physics  and Astronomy, Texas A$\&$M University,
College Station, Texas 77843, USA}\\
\normalsize\emph{$^{\bf 2}$Department of Physics and Astronomy, University of Hawaii, Honolulu, Hawaii 96822, USA }\\
\vspace{1.0cm}
}

\begin{abstract}
We consider the effect on early Universe cosmology of the dark photon associated with the
gauging of $U(1)_{T3R}$, a symmetry group under which only right-handed Standard Model fermions
transform non-trivially.  We find that cosmological constraints on this scenario are qualitatively
much more severe than on other well-studied cases of a new $U(1)$ gauge group, because the dark photon
couples to chiral fermions.  In particular, the dark photon of $U(1)_{T3R}$ is always produced and
equilibrates in the early Universe, no matter how small the gauge coupling, unless the symmetry-breaking
scale is extremely large.  This occurs because, no matter how the weak the coupling, the Goldstone mode
(equivalently, the longitudinal polarization) does not decouple.  As a result, even the limit of an extremely
light and weakly-coupled dark photon of $U(1)_{T3R}$ is effectively ruled out by cosmological constraints,
unless the symmetry-breaking
scale is extremely large. We also discuss the possibility of ameliorating the Hubble tension in this model.
\end{abstract}
\maketitle


\section{Introduction}

In recent times, precise measurements of the cosmic microwave background (CMB) by the Planck experiment have
placed tight constraints on the number of effective relativistic degrees of freedom
in the early universe, encoded in the quantity $\Delta N_{eff}$~\cite{Aghanim:2018eyx}.
These constraints can rule out models of new physics with
new low-mass particles.  Recent work has considered the constraints imposed on
models of new physics in which a low-mass dark photon ($A'$) couples to Standard Model
(SM) (see, for example,~\cite{Kamada:2015era,Kamada:2018zxi,Knapen:2017xzo,Escudero:2019gzq,Sabti:2019mhn,Foot:2014uba}).
These works have focused on scenarios in which the dark photon is either secluded
(coupling to SM particles only via kinetic mixing) or couples to the charges $B-L$  or
$L_i - L_j$~\cite{Foot:1990mn,He:1990pn,He:1991qd}.  But another well-studied anomaly-free choice of new $U(1)$ gauge group
is $U(1)_{T3R}$; in this scenario, only one or more complete generations of right-handed
SM fermions are charged, with up-type and down-type fermions having opposite charge.  This
scenario was originally considered in the context of left-right models~\cite{Pati:1974yy,Mohapatra:1974gc,Senjanovic:1975rk},
in which $U(1)_{T3R}$
is the diagonal subgroup of $SU(2)_R$, under which right-handed fermions transform as
doublets. Recently, $U(1)_{T3R}$ has been investigated for the purpose of building a well-motivated model of sub-GeV dark matter~\cite{Dutta:2019fxn}. This model explains the hierarchies among the light fermion masses and contains a light gauge boson and a light scalar particle. 
In this brief letter, we point out that the dark photon of $U(1)_{T3R}$ contributes to
$\Delta N_{eff}$ in a manner which is qualitatively different than the dark photon of other
well-studied examples, such as $B-L$, $L_i-L_j$, a secluded $U(1)$, etc.

In these other well-studied examples, there are generally two ways in which one can ensure that
the contribution of the dark photon to $\Delta N_{eff}$ is negligible; either the dark photon
can be heavy enough that its abundance is negligible due to Boltzmann suppression
at the time of neutrino decoupling, or its
coupling can be so weak that it is never produced in the early Universe, again leading to a negligible
abundance.  But if the dark photon is the gauge boson of $U(1)_{T3R}$, then this second option is
foreclosed; {\it the dark photon is always produced in the early Universe, no matter how weak the coupling}
unless the symmetry-breaking scale is $\gtrsim 10^6 \gev$.

This result might at first seem counterintuitive.  But one way to see this is to note that
$m_{A'} \propto g V$, where $g$ is the gauge coupling and $V$ is the expectation value of the
field which breaks the $U(1)$ gauge symmetry.  Thus, for fixed $V$, as the gauge coupling gets
weaker, the mass of the dark photon becomes smaller.  If we then consider an inverse decay
process like $\bar f f \rightarrow \gamma A'$, the sum over $A'$ polarizations yields a factor
of $-(g^{\mu \nu} - k^\mu k^\nu / m_{A'}^2)$, where the second term arises due to contribution
of the longitudinal polarization.  The $m_{A'}^2$ factor in the denominator cancels the $g^2$ factor
in the squared matrix element, potentially leaving a finite term even at arbitrarily small coupling.

Of course, these considerations apply for any choice of $U(1)$.  For any choice of the $U(1)$ gauge
group, $m_{A'} \propto gV$, and the longitudinal polarization thus always receives an enhancement
which is proportional to $1/g$.  But the enhanced term in the polarization sum is $\propto k^\mu k^\nu$;
in cases where $A'$ couples to SM fermions through a purely
vector interaction,  the resulting term in the matrix element is zero due to the Ward Identity.
However, if the gauge group is $U(1)_{T3R}$, then the longitudinal polarization is contracted
with a combination of vector and axial vector currents, and the axial vector term does not vanish. This feature causes the $A'$ production from the SM fermions to be nonzero even in the limit where the $A'$ gauge coupling 
is taken to be very small.

Another way to see this result is to note that, in the weakly coupled limit, the $U(1)$ gauge group
essentially becomes a global symmetry group, and the transverse polarizations of the $A'$ manifestly
decouple.  But the longitudinal polarization instead becomes the massless Goldstone mode of the
spontaneously broken global symmetry, which need not decouple.  Again, these considerations apply for
any choice of the $U(1)$ gauge group.  But the relevant question is how does the Goldstone mode
couple to SM fermions.  The coupling of the Goldstone mode derives from the complex scalar whose vacuum expectation value 
(vev) breaks the $U(1)$ symmetry; the Goldstone is the real excitation orthogonal to direction of the
symmetry breaking vev.  Since an unbroken $U(1)_{T3R}$ would forbid a SM fermion mass, the coupling
of the Goldstone boson to any SM fermion charged under $U(1)_{T3R}$ must scale as $m_f/V$.  But if
the dark photon instead couples to $B-L$ or $L_i - L_j$, there is no reason why the symmetry-breaking
field need have a sizeable coupling to SM fields at the era of neutrino decoupling.   As a result of
these considerations, we will find that the scenario in which  $U(1)_{T3R}$ is gauged is much more
tightly constrained by cosmological observations than other recently studied scenarios. We will see explicitly that these stringent constraints  emerge when the relativistic new gauge bosons are produced directly 
from on-shell muons. As a result, 
if only second-generation fermions are charged under 
$U(1)_{T3R}$, then collider and other astrophysical constraints are largely unaffected by these considerations, whereas constraints 
arising from early Universe cosmology become much more 
severe.  


\section{Production of $A'$ in the Early Universe}

For simplicity, we assume that only second generation right-handed fermions are charged under
$U(1)_{T3R}$, with up-type and down-type fermions having opposite charge ($Q_{c_R} =
Q_{\nu_R}=1$, $Q_{s_R} = Q_{\mu_R}=-1$).  One can verify that this choice is anomaly-free.
We will assume that $g\ll 1$, where $g$ is the coupling of $U(1)_{T3R}$.  In that case,
the dominant processes by which $A'$ can be produced in the early Universe are inverse decay
processes, in which only one factor of $g$ is appears in the matrix element.  In~\cite{Escudero:2019gzq},
it was argued that the dominant production process is $\bar \mu \mu \rightarrow \gamma A'$.  For our
purpose, it will be sufficient to consider this process in order to demonstrate that $A'$ is always
produced in the early Universe, provided this process is kinematically allowed and the symmetry-breaking
scale is not extremely large.

The relevant Lagrangian for the gauge boson $A'$ is
\bea
{\cal L} &=& -\frac{1}{4}B_{\mu \nu} B^{\mu \nu} + \imath g (\phi \partial_\mu \phi^* - \phi^* \partial_\mu \phi) {A'}^\mu
+g^2 \phi \phi^* A'_\mu {A'}^\mu
-g \sum_f Q_f \bar f_R \gamma_\mu f_R {A'}^\mu,
\eea
where $B_{\mu \nu} = \partial_\mu A'_\nu - \partial_\nu A'_\mu$,
$\phi$ is complex scalar field charged under $U(1)_{T3R}$, and $\langle \phi \rangle =V$.
The condensation of $\phi$ spontaneously breaks $U(1)_{T3R}$, giving the dark photon a mass
$m_{A'}^2 = 2 g^2 V^2$.

We may express  the excitation of $\phi$ about its vev in terms of two real fields,
$\phi'$ and $\phi_I$, yielding
$\phi = V + (1\sqrt{2}) \phi' + (\imath\sqrt{2}) \phi_I$.
$\phi'$ is the dark Higgs, and is a physical real scalar excitation.
$\phi_I$ is the Goldstone mode, which is absorbed by dark photon in order to
provide the third physical polarization of the $A'$.

The matrix element for the process $\bar f(p_2) f(p_1) \rightarrow \gamma(k_2) A'(k_1)$ is
given by
\bea
\imath {\cal M}_{A'} &=&
-\imath \frac{e Q_f m_{A'}}{\sqrt2 V} \bar v(p_2) \left[
\frac{\gamma^\mu \Dsl k_2 \gamma^\nu - 2 p_2^\mu \gamma^\nu }{-2p_2 \cdot k_2 +k_2^2}
+ \frac{2p_1^\mu \gamma^\nu  - \gamma^\nu \Dsl k_2  \gamma^\mu }{-2p_1 \cdot k_2 +k_2^2 }
\right] \frac{1 + \gamma^5}{2} u (p_1) \epsilon^*_\nu (k_1) \epsilon^*_\mu (k_2),
\eea
where $\epsilon(k_1)$ ad $\epsilon(k_2)$ are the polarization vectors of the
$A'$ and $\gamma$, respectively.  The $P_R = (1+\gamma^5)/2$ projector appears
because $A'$ only couples to $f_R$.

One can easily verify that the matrix element
vanishes under the replacement $\epsilon^\mu (k_2) \rightarrow k_2^\mu$, as required
by the Ward Identity.  But one can also verify that, under
the replacement $\epsilon^\nu (k_1) \rightarrow k_1^\nu$, the only non-vanishing term
is proportional the one proportional to $\gamma^5$.  This is also a result of the
Ward Identity.  If the $\gamma^5$ term had been removed, then the coupling of $f$ to
$A'$ would have been a pure vector interaction, and contracting the external momentum
into the vector current necessarily yields zero.

This result immediately indicates that, in the case where the $A'$ coupling to SM fermions
is a pure vector interaction, the longitudinal polarization yields no parametric enhancement
to the matrix element.  The squared matrix element is contracted with an $A'$ polarization sum
factor given by $- (g^{\mu \nu} - k_1^\mu k_1^\nu /m_{A'}^2)$.  In the weak coupling limit
($m_{A'}/V \propto g \rightarrow 0$), the second term receives a parametric enhancement, but vanishes
identically when contracted into a purely vector current.

We are interested in squared matrix element in limit where $g \ll 1$.  In this case,
only the $k_1^\mu k_1^\nu /m_{A'}^2$ term in the polarization sum  is relevant, as this
is the only term which can yield a non-zero contribution which contracted with a matrix
element that scales as $g^2$.  From the Ward Identity, we see that we need only consider
the term in the matrix element proportional to $\gamma^5$.  Summing over the polarizations of
the $A'$, we thus find
\bea
\sum_{A'~pols} |{\cal M}_{A'}|^2 &=&
\left(\frac{e m_f} {2\sqrt{2} V} \right)^2
\left| \bar v(p_2)
\left[
\frac{\gamma^\mu \Dsl k_1  }{p_2 \cdot k_2 }
- \frac{2 p_1^\mu  - \Dsl k_2 \gamma^\mu }{p_1 \cdot k_2 }
\right] \gamma^5 u (p_1)  \epsilon^*_\mu (k_2) \right|^2 ,
\eea
where we have set $k_2^2=0$ and $Q_f=-1$.  It is thus clear that a finite piece is left, even in the
limit $g\rightarrow 0$, when the dark photon couples to a chiral fermion.

One can verify this result straightforwardly by considering the limit where $g=0$, in which
case the $A'$ is exactly massless, and $U(1)_{T3R}$ becomes effectively a global symmetry.
In this case, the transverse polarizations of the $A'$ must decouple, but the coupling of
the massless Goldstone mode should reproduce the above squared matrix element.  Indeed,
this intuition is easily verified.  The coupling of the Goldstone mode to $f$ is induced from
the coupling of the symmetry-breaking field $\phi$ to $f$, which is required in order for
the fermion mass to be generated from a gauge-invariant Yukawa coupling.  In the effective
field theory defined below the electroweak symmetry breaking scale, we find
\bea
{\cal L}_{yuk.} &=& \lambda_f \phi \bar f  \left(\frac{1 + \gamma^5}{2} \right) f
+ \lambda_f \phi^* \bar f  \left(\frac{1 - \gamma^5}{2} \right) f
\nonumber\\
&=&
m_f \bar f f
+ \frac{m_f}{\sqrt{2}V} \phi' \bar f  f + \imath \frac{m_f}{\sqrt{2}V} \phi_I \bar f \gamma^5 f ,
\eea
implying that the Goldstone mode $\phi_I$ couples to $f$ as a pseudoscalar with coupling
$m_f / \sqrt{2} V$.

It is then straightforward to compute the squared matrix element for the process
$\bar f(p_2) f(p_1) \rightarrow \gamma (k_2) \phi_I(k_1)$, yielding
\bea
|{\cal M}_{Gold.}|^2 &=&
\left( \frac{e m_f}{2\sqrt{2} V} \right)^2
\left| \bar v(p_2)
\left[
\frac{ \gamma^\mu  \Dsl k_1}{p_1 \cdot k_1 }
-  \frac{2 p_1^\mu - \Dsl k_2 \gamma^\mu }{p_1 \cdot k_2 }
\right]
\gamma^5 u (p_1)  \epsilon^*_\mu (k_2) \right|^2 .
\eea
In the limit $m_{A'} =0$, we find $p_1 \cdot k_1 = p_2 \cdot k_2$, implying that the cross section
for producing the massless $A'$ in the weakly coupled limit is equal to the cross section for producing
the massless Goldstone boson, as required by the Goldstone Equivalence Theorem.

From here on, it is convenient to proceed in the Goldstone limit, where we take $g=0$.  If we choose
simple kinematics for the incoming SM fermions, $p_1^\mu = (E, \vec{p})$,
$p_2^\mu = (E, -\vec{p})$, defining $p = |\vec{p}|$, we find
\bea
\sigma v &=& \frac{\alpha_{em} m_f^2}{4 E^2 V^2}
\left[\frac{(2E^2 + p^2) \tanh^{-1} (p/E)}{E p} -1 \right] .
\eea
As expected, the cross section scales as $\alpha m_f^2 / V^2$, since the coupling of
the Goldstone mode to $f$ is inherited from the coupling of the symmetry-breaking field,
which necessarily scales as $m_f/V$, since $U(1)_{T3R}$ protects the fermion mass.
We find that the thermally averaged cross section is given by 
\bea
\langle \sigma v \rangle_{T \sim m_f} \sim 0.18 \frac{\alpha_{em} }{V^2} ,
\eea

To determine the range of $V$ for which $A'$ equilibrates in the early Universe, we explicitly solve the 
Boltzmann equation for the $A'$ abundance.  But following~\cite{Escudero:2019gzq}, we find an approximate   
criterion for $A'$ to not have equilibrated in the early Universe:
\bea
\eta_{f,\bar f} (T=m_f) \langle \sigma v \rangle_{T \sim m_f} &\lesssim& H = \sqrt{\frac{ g_* \rho_{rad} (T=m_f)}{3 M_{pl}^2}} ,
\eea
where $M_{pl}$ is the reduced Planck mass and $g_*$ is the effective number 
of Standard Model relativistic degrees of freedom at $T=m_f$, yielding
\bea
\langle \sigma v \rangle_{T \sim m_f} &\lesssim & \frac{2.2 \sqrt{g_*}}{2 m_f M_{pl}} .
\eea
We then find that $A'$ will have equilibrated in the early Universe unless
\bea
V &\gtrsim&  \left(\frac{0.18}{2.2} \frac{2 \alpha_{em} m_f M_{pl}}{\sqrt{g_*}} \right)^{1/2} ,
\nonumber\\
&\gtrsim& ( 9 \times 10^6~\gev) \left[ \left(\frac{m_f}{m_\mu} \right)
\left(\frac{g_*}{16.02} \right)^{-1/2} \left(\frac{\alpha_{em}}{1/137} \right)\right]^{1/2} .
\eea
A solution of the Boltzmann equation yields a similar result.


\section{$\Delta N_{eff}$}

Given that $A'$ is produced and equilibrates in the early Universe, we must now determine
how its abundance at the time of recombination corrects $N_{eff}$.  For this purpose, we
will assume that the neutrino mixing angle is small (the sterile neutrino mass eigenstate,
$\nu_s$, is almost entirely $\nu_R$), and that $m_{\nu_S} >10\mev$.
If $m_{A'} > 10~\mev$, then the $A'$ abundance is heavily Boltzmann-suppressed at the time
of neutrino decoupling, and its impact on $N_{eff}$ is negligible~\cite{Escudero:2019gzq}.

In the limit $m_{A'} / V \rightarrow 0$, the transverse polarizations of the $A'$ completely
decouple, and we are left with a massless Goldstone mode, which thermalizes in the early Universe and
decouples before neutrino
decoupling, and which does not decay.  As a result, the Goldstone degree of freedom is at the same temperature
as the neutrinos, and its energy density at recombination contributes as $\Delta N_{eff} = 4/7$.

If $m_{A'}$ is non-negligible, but $m_{A'} < 1\mev$, then the $A'$ can decay to $\nu_A \nu_A$ through
a one-loop process (decay to $\gamma \gamma$ is forbidden by the Landau-Yang Theorem).  As the temperature
drops well below $m_{A'}$, $A'$ decays will heat the neutrino population, leading to an
even larger value of $\Delta N_{eff}$~\cite{Escudero:2019gzq,Sabti:2019mhn}.

But if $m_{A'}$ lies in the range $\sim 1-10\mev$. the analysis is model-dependent.  In particular, $A'$
can also decay to $e^+ e^-$ through a one-loop kinetic-mixing process.
The relative branching fractions for
$A'$ decay to $\nu_A \nu_A$ and $e^+ e^-$ are determined by the details of the neutrino mass matrix.
This yields two relevant effects. First, electrons and neutrinos can remained coupled via decays and
inverse decays of $A'$, delaying the time neutrino decoupling.  As shown in~\cite{Escudero:2019gzq}, this
can yield an ${\cal O}(1)$ correction to the allowed mass range for $m_{A'}$.  But an even more significant
effect arises if the branching fraction for $A' \rightarrow e^+ e^-$ can be large.
If the
dominant decay of $A'$ is to $\nu_A \nu_A$, then little changes from the above analysis.  But if the
dominant decay of $A'$ is to $e^+ e^-$, then when the temperature drops well below $m_{A'}$, the photon
temperature increases, yielding a negative contribution to $\Delta N_{eff}$.  With an appropriate choice
of branching fraction, $\Delta N_{eff}$ can be tuned to be arbitrarily small.

\begin{figure}[h]
\begin{subfigure}[b]{0.485\textwidth}
\includegraphics[width=1.0\linewidth,height=6cm]{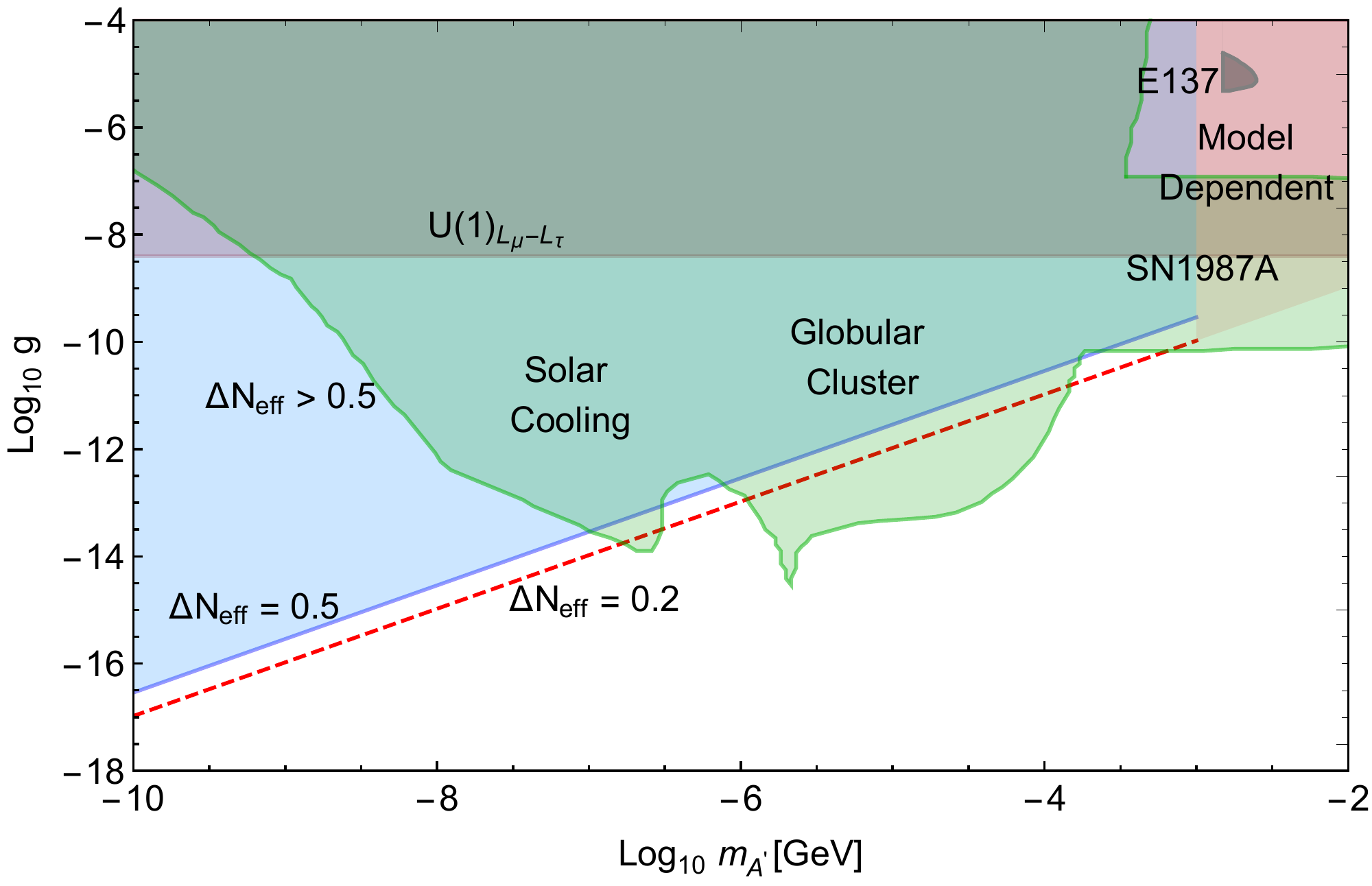}
\captionsetup{labelfont = bf}
\caption{\label{fig:Neff:gvsmA}}
\end{subfigure}	
\hspace{0.2cm}	
\begin{subfigure}[b]{0.485\textwidth}
\includegraphics[width=1.0\linewidth,height=6cm]{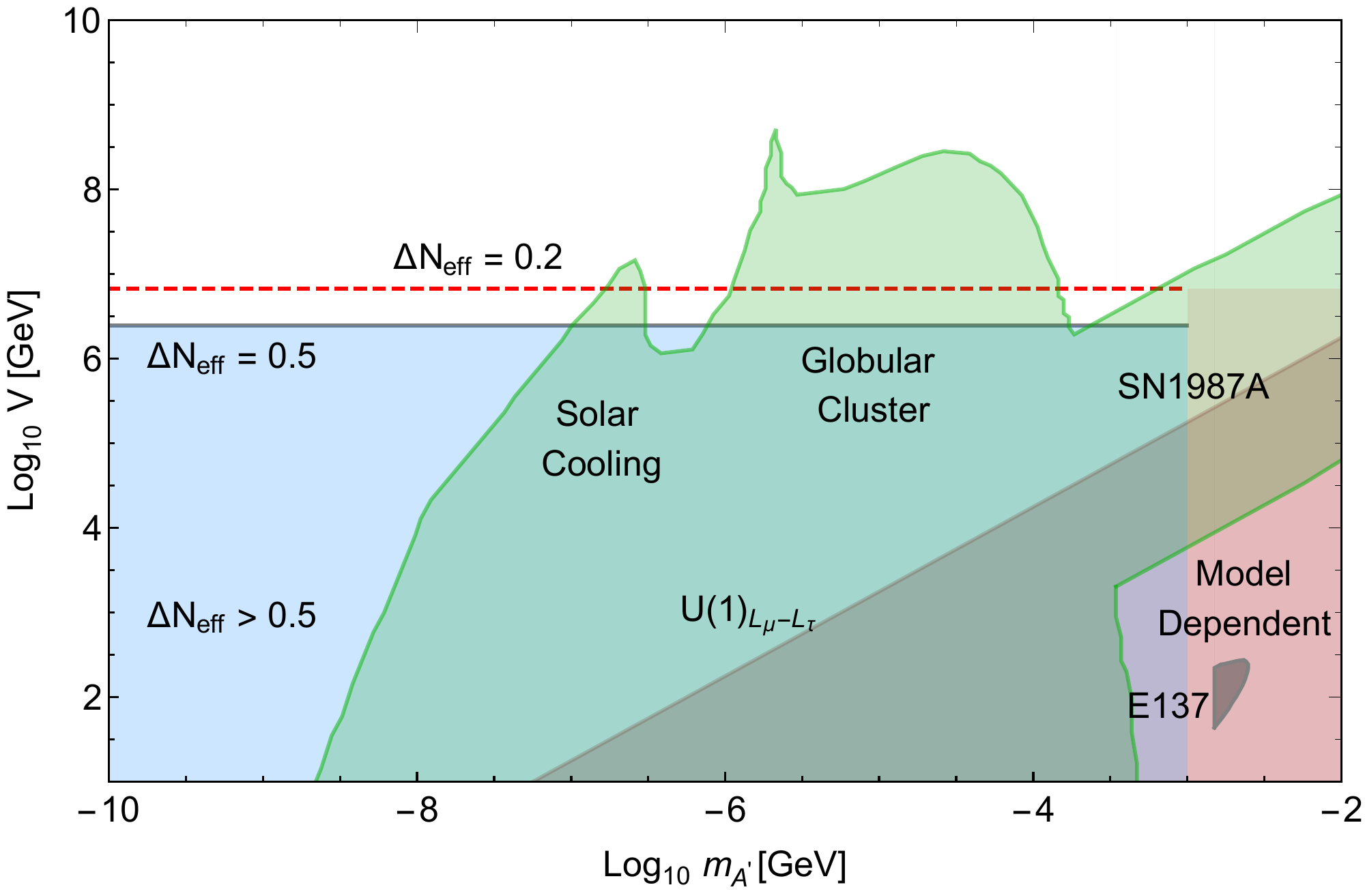}	
\captionsetup{labelfont = bf}
\caption{\label{fig:Neff:VvsmA}}
\end{subfigure}	
\captionsetup{justification   = RaggedRight,
             labelfont = bf}
\caption{\label{fig:Neff} The excluded regions of parameter space ($\Delta N_{eff} \geq 0.5$) in the $(m_{A'}, g)$-plane (left panel) and $(m_{A'}, V)$-plane (right panel).  In blue is the excluded region if $A'$ couples to $U(1)_{T3R}$, under which second generation Standard Model fermions are charged.  For the case where the $A'$ couples to $L_\mu - L_\tau$, the excluded region in purple is reproduced from~\cite{Escudero:2019gzq}.  In both cases, the range $1~\mev \leq m_{A'} \leq 10~\mev$ is shaded, as exclusion contours in this mass range depend on details of the model.} 
\end{figure}

In \textbf{Fig.~\ref{fig:Neff:gvsmA}} we plot the excluded region of parameter space in the $(m_{A'}, g)$-plane for the case where $A'$ couples to $U(1)_{T3R}$ (blue), along with similar results from~\cite{Escudero:2019gzq} (purple) for the case where $A'$ couples to $L_\mu - L_\tau$.
Note, the $A'$ abundance produced via inverse decay is 
computed by solving the Boltzmann equation.
To facilitate comparison with~\cite{Escudero:2019gzq}, we will treat as excluded models for which
$\Delta N_{eff} \geq 0.5$.
The red dashed line indicates the parameter 
space for which $A'$ will not fully equilibrate, yielding 
$\Delta N_{eff} \sim 0.2$.
In \textbf{Fig.~\ref{fig:Neff:VvsmA}},
we plot the excluded regions of parameter space in the $(m_{A'}, V)$-plane. The larger V values correspond to regions where $A'$ will not be in equilibrium.

In both \textbf{Fig.~\ref{fig:Neff:gvsmA}} and \textbf{Fig.~\ref{fig:Neff:VvsmA}}, we show the excluded region of parameter space constrained by  relevant fixed target experiments and different astrophysical processes. In electron beam dump experiments, such as SLAC E137~\cite{Riordan:1987aw, Bjorken:1988as, Bjorken:2009mm, Andreas:2012mt}, the $A^\prime$ can be produced via $e$-bremsstrahlung and decay to a $e^+ e^-$ pair through loop-suppressed mixing with photon. Due to the loop-suppression, other fixed target experiments are irrelevant in the parameter space of our interest~\cite{Bauer:2018onh}. Note, E137 can only provide bounds in the model dependent region i.e. $m_{A^\prime} > 1 $~MeV. The $A^\prime$ can be produced inside the core of a supernova through the mixing with the photon, and  
can subsequently escape, resulting in energy loss. Constraints on this process are found by observing the energy loss of SN1987A~\cite{Dent:2012mx, Harnik:2012ni}. The $A^\prime$ 
can also be produced in the Sun and can contribute to the solar cooling process. By requiring that the luminosity due to the dark photon be sufficiently small compared to the luminosity due to the photon, bounds can be derived~\cite{Redondo:2008aa, Harnik:2012ni}. Bounds can be found from the cooling of stars in Globular clusters in a similar way~\cite{Harnik:2012ni}. The green region in \textbf{Fig.~\ref{fig:Neff:gvsmA}} and \textbf{Fig.~\ref{fig:Neff:VvsmA}} shows the combined excluded region, considering the cooling of supernovae, the Sun and Globular clusters. Note, the parameter space for the lower mass 
range is not constrained by the astrophysical constraints, 
but is tightly constrained by the cosmological bounds we have 
found.    

Note, the longitudinal polarization of the 
$A'$ has negligible effect on the collider and astrophysical 
bounds.  The reason is because, in all of those cases, the 
$A'$ is produced through kinetic mixing with the photon, 
and its longitudinal polarization necessarily decouples.  The 
more stringent constraints on $U(1)_{T3R}$ which arise from 
production of the longitudinal polarization come into play 
only when the relativistic $A'$s are produced directly 
from on-shell muons.  As a result, these cosmological 
constraints are uniquely constraining.

It has been noted (see, for example,~\cite{Escudero:2019gzq,Bernal:2016gxb, Alcaniz:2019kah,Vagnozzi:2019ezj}) that the tension 
between the determination of $H_0$ from low-$z$ measurements~\cite{Riess:2016jrr,Riess:2018byc} and from the CMB~\cite{Aghanim:2018eyx}
can potentially be resolved if $\Delta N_{eff} \sim 0.2-0.5$. This range of $\Delta N_{eff}$ can arise in this model for a large range of $m_{A'}$.  In Figure \ref{fig:Neff}, we show the parameter space where $\Delta N_{eff} \sim 0.2-0.5$ by solving the Boltzmann equation. In the model dependent part of the parameter space, $m_{A'}\sim 1-10$ MeV, $\Delta N_{eff}$ can be set to  $\sim 0.2-0.5$ by appropriately choosing the mixing between the active and sterile component which determines the branching fraction for
$A'$ decay to $\nu_A \nu_A$. In this case, $\Delta N_{eff}$ can receive both
positive and negative contributions which can be tuned against each other by tuning the branching fraction for
$A'$ decay to $\nu_A \nu_A$ and $e^+ e^-$. For $m_{A'} \gtrsim 10$ MeV, we choose
the gauge coupling appropriately to obtain the correct $\Delta N_{eff}$.


\section{Conclusion}

We have considered the effect of the dark photon of $U(1)_{T3R}$ on cosmology in
the early Universe.  We have found that, unlike other recently studied cases, such
as $B-L$ and $L_i-L_j$, if the dark photon is the gauge boson of $U(1)_{T3R}$, cosmological
constraints are much tighter.  In particular, $A'$ is always produced and equilibrates in
the early Universe, not matter how small the gauge coupling is, provided the symmetry
breaking scale is $\lesssim 10^6 \gev$ (for the case where second generation right-handed fermions
are charged under $U(1)_{T3R}$).  Even if the gauge coupling is made arbitrarily small, this
suppression of the $A'$ production cross section is compensated by the enhancement of the
longitudinal polarization when there is an axial vector coupling.  This amounts to saying that,
even in the limit when coupling becomes negligible and the symmetry becomes global, the
Goldstone mode remains coupled to the charged fermions.  We calculated $\Delta N_{eff}$ from the $A'$ abundance by solving Boltzmann equation for this model and showed contours of $\Delta N_{eff}=0.2$, 0.5 along with various constraints, e.g., collider, beam dump,  cooling of supernova, Sun and Globular clusters etc. We found that the cosmological constraints obtained  in this work can 
exclude a large region of parameter space which is allowed 
by all other laboratory or astrophysical constraints.

We could consider the same scenario in the case where right-handed first generation fermions
are instead charged under $U(1)_{T3R}$.  The considerations described above are largely
unchanged; in this case, $A'$ is produced and equilibrates in the early Universe unless
the symmetry-breaking scale is $> 10^5 \gev$.  One difference occurs if $m_{A'}$ lies in the
$1-10 \mev$ range.  In this case, assuming the sterile neutrino is heavy, one finds that the
$A' \rightarrow \nu_A \nu_A$ decay process is one-loop suppressed, while $A' \rightarrow e^+ e^-$ decay
occurs at tree-level.  Thus, one would generally expect $A'$ decay to inject energy into the
photon gas, yielding a negative contribution to $N_{eff}$.

We see that regions of parameter space at very small $m_{A'}$ found in~\cite{Dutta:2019fxn}
are in fact in tension with cosmological constraints.  In particular, this would rule out the scenarios
described in~\cite{Dutta:2019fxn} in which the dark photon coupled to electrons.  Models in which $m_{A'} > 10\mev$
are still consistent with cosmological constraints, but if $A'$ couples to right-handed electrons, then they
are in tension with atomic parity violation experiments.  But it may be possible to relax the tension with
atomic parity violation experiments with a modest fine-tuning against additional sources of new physics; it
would be interesting to investigate this further.

We also discussed the possibilities of ameliorating Hubble parameter measurements in this model which requires $\Delta N_{eff} \sim 0.2-0.5$. This range of $\Delta N_{eff}$ can arise in this model for a large range of $m_{A'}$.  We showed that for $m_{A'} <1$ MeV, some parts of the required $\Delta N_{eff}$ range are allowed by all other astrophysical constraints. In the model dependent part of the parameter space, $m_{A'}\sim 1-10$ MeV, $\Delta N_{eff}$ can be set to  $\sim 0.2-0.5$ by appropriately choosing the mixing between the active and sterile component and for $m_{A'} \gtrsim 10$ MeV, the gauge coupling can be appropriately chosen to obtain the required$\Delta N_{eff}$.

We have focused in particular on the case where $U(1)_{T3R}$ is gauged.  But the general result is valid in
any scenario in which the dark photon has a chiral coupling to SM fermions.  One would expect any such model
to be tightly constrained by early Universe cosmology.

{\bf Acknowledgments}

The work of BD and SG are supported in part by the DOE Grant No.~DE-SC0010813.
The work of JK is supported in part by DOE Grant No.~DE-SC0010504.

\bibliographystyle{apsrev4-1.bst}
\bibliography{Neff}

\end{document}